\documentclass[11pt]{article}

\usepackage[final]{acl}
\usepackage{times}
\usepackage{enumitem}
\usepackage{latexsym}
\usepackage[T1]{fontenc}
\usepackage[utf8]{inputenc}
\usepackage{microtype}
\usepackage{inconsolata}
\usepackage{graphicx}
\usepackage{placeins}
\usepackage{hyperref}       
\usepackage{url}            
\usepackage{booktabs}       
\usepackage{amsfonts}       
\usepackage{nicefrac}       
\usepackage{xcolor}         
\usepackage{amsmath}
\usepackage{multirow}
\usepackage{pifont}
\usepackage{tikz}
\usetikzlibrary{arrows.meta}
\newsavebox{\findingboxbox}
\newenvironment{findingbox}{%
\par\smallskip
\begin{lrbox}{\findingboxbox}
\begin{minipage}{0.94\linewidth}
\small
}{%
\end{minipage}
\end{lrbox}
\noindent\fcolorbox{gray!55}{gray!8}{\usebox{\findingboxbox}}
\par\smallskip
}

\newcommand{\benchmarkname}{PERFOPT-Bench}
\newcommand{\numtasks}{12}
\newcommand{\numconfigs}{7}
\newcommand{\cmark}{\ding{51}}
\newcommand{\xmark}{\ding{55}}

\title{PERFOPT-Bench: Evaluating Coding Agents on Software Performance Optimization}

\author{
Yingyun Cui \thanks{Equal contribution.} \\
OPPO Research Institute \\
\texttt{cuiyingyun@oppo.com}
\And
Yi Xie \footnotemark[1] \\
University of Arizona \\
\texttt{yix@arizona.edu}
\And
Piaohong Wang \footnotemark[1] \\
OPPO Research Institute \\
\texttt{wangpiaohong@oppo.com}
\AND
Jiawei Ma\thanks{All these authors are corresponding authors.}  \\
Department of Computer Science \\ \& Institute of Digital Medicine, \\City University of Hong Kong  \\
\texttt{jiaweima@cityu.edu.hk}
\And
Bo Liu\footnotemark[2]  \\
University of Arizona \\
\texttt{boliu@arizona.edu}
\And
Liangliang Cao\footnotemark[2] \\
The Hong Kong \\ Polytechnic University \\
\texttt{liangliang.cao@gmail.com}
}

\usepackage[most]{tcolorbox}
\usepackage{xcolor}
\begin{document}
\maketitle

\begin{abstract}
Coding-agent benchmarks have largely measured whether agents can produce
functionally correct patches, but production software also demands
measurable speedups on real execution targets. Performance optimization is
a distinct agentic task: agents must profile executions, diagnose
cross-layer bottlenecks, edit code without breaking correctness, and verify
that gains are reproducible rather than measurement artifacts.
We introduce \benchmarkname{}, a benchmark for evaluating this full
performance-engineering loop. Each task provides a correct but deliberately
suboptimal codebase and asks the agent to improve a target performance
metric; scoring requires hidden correctness tests, verified-speedup
measurement, and trajectory-level audit.
We evaluate \numconfigs{} agent stacks with different LLMs and agent
frameworks on \numtasks{} long-horizon optimization tasks. The results show
that optimization performance is workload-dependent rather than determined
by model identity alone: no single stack dominates, and changing the agent
framework can materially change the same LLM's per-task speedup profile. We further find
that raw speedup is unsafe as a benchmark score, since some large gains
arise from benchmark-specific shortcut exploitation; an exploratory relay
pilot suggests that restarting from an externalized optimization summary
can recover additional headroom after an initial session stops.
The benchmark and our evaluation are available at:
\url{https://anonymous.4open.science/r/Dataset-D3CC}.
\end{abstract}

\section{Introduction}
\label{sec:intro}

Coding agents increasingly operate as autonomous
developers: inspecting files, executing tests, and revising code based
on feedback~\citep{jimenez2024swebench,yang2024sweagent,
wang2025openhands,merrill2026terminalbench}. Benchmarks have followed
this shift, moving from function-level synthesis~\citep{
chen2021evaluating,austin2021program} to repository repair and
terminal environments~\citep{jimenez2024swebench,jain2024livecodebench,
zhuo2025bigcodebench,merrill2026terminalbench}. Yet most evaluations
still credit agents for passing tests or resolving issues, leaving
performance largely outside the scoring target. This omission matters
in production settings, where code must remain efficient and stable
across hardware, runtime environments, and resource
constraints~\citep{williams2009roofline,nickolls2008scalable,
dean2013tail}.

Software performance tuning differs fundamentally from writing a
passing patch. Source-code inspection alone is insufficient; agents
must profile real executions, diagnose system-level bottlenecks, and
account for platform dependence. A change that accelerates one
architecture can slow another, so the central evaluation question
shifts from \emph{Can the agent produce correct code?} to \emph{Can it measurably optimize a system on the target hardware?} This shift
surfaces challenges that single-shot code generation does not.
An agent must
(i)~design and run a profile--patch--verify inner loop without
external scaffolding;
(ii)~hold the optimization objective steady across very long
execution trajectories;
(iii)~organize and compress measurement history when reasoning spans
multiple context windows; and
(iv)~carry enough systems knowledge to recover from diverse failure
modes such as bottleneck misidentification, ineffective
micro-optimization, correctness regression, and unstable measurement.
Prior work covers parts of this picture -- algorithmic
optimization~\citep{shypula2024learning}, resource-efficient
generation~\citep{du2024mercury,huang2024effibench}, repository
performance repair~\citep{he2025sweperf,ma2025swefficiency}, and GPU
operator synthesis~\citep{ouyang2025kernelbench,li2025tritonbench} --
but none require an agent to complete the full
performance-engineering loop autonomously: navigating a codebase,
profiling the target hardware, modifying implementations without
breaking correctness, and empirically validating the gains.

We introduce \benchmarkname{} to evaluate that loop directly. Each
task pairs a functionally correct but suboptimal codebase with a
natural-language optimization objective. Agents must improve a target
metric without breaking correctness, while interacting with the
system through profiling, editing, and empirical feedback. On this benchmark, 
we evaluate \numconfigs{} agent stacks
across \numtasks{} long-horizon tasks. We find that
optimization capability is workload-dependent and not a property of
the LLM alone: the best stack changes across tasks, and agent frameworks
shape the same LLM's per-task speedup profile.
We also find that agent relay may expose additional optimization
headroom. Our contributions are threefold:
\begin{itemize}[leftmargin=*,nosep,topsep=2pt]
\item \benchmarkname{}, a cross-layer software performance optimization
benchmark with verified speedup metrics and hidden correctness checks,
covering a wide range of optimization problems, including toolchain configuration, SIMD usage,
memory locality, cache behavior, and algorithmic restructuring.

\item A semi-automated construction pipeline that separates task authoring
from evaluated agents, combining LLM-driven category generation, private
reference solving, pilot-solver calibration, and expert curation to produce
diverse performance-optimization tasks.

\item An empirical study of seven agent stacks on twelve tasks, showing that
optimization performance is workload-dependent, so that agent frameworks can
materially change the behavior of the same LLM, and that---because
performance is environment-conditioned---raw speedups must be read
case-by-case and validated by combined executable checks and expert audit
rather than by passing tests alone; we additionally report an exploratory one-step relay
pilot as a continuation stress test.
\end{itemize}
\section{Benchmark Construction}

\providecommand{\hobFull}{\textcolor{green!50!black}{\cmark}}
\providecommand{\hobPart}{\textcolor{orange!80!black}{\ensuremath{\triangle}}}
\providecommand{\hobNo}{\textcolor{red!65!black}{\xmark}}

\begin{table*}[t]
\centering
\small
\setlength{\tabcolsep}{2.6pt}
\renewcommand{\arraystretch}{0.96}

\caption{Comparison with related coding-agent and performance
benchmarks. \benchmarkname{} targets the intersection of these eight
dimensions; abbreviations are defined below.}
\label{tab:benchmark-comparison}

\resizebox{\textwidth}{!}{%
\begin{tabular}{@{}llcccccccc@{}}
\toprule
\textbf{Benchmark}
& \textbf{Task type}
& \textbf{\shortstack{Perf.\\Opt.}}
& \textbf{\shortstack{Cross-\\Layer}}
& \textbf{\shortstack{Cont.\\Score}}
& \textbf{\shortstack{Real\\HW}}
& \textbf{\shortstack{Bottleneck\\Dx}}
& \textbf{\shortstack{Auto-\\Gen.}}
& \textbf{\shortstack{Real\\Repo}}
& \textbf{Agentic} \\
\midrule

\multicolumn{10}{@{}l}{\textit{Function / algorithm level}} \\
HumanEval\cite{chen2021evaluating}
& Function generation
& \hobNo & \hobNo & \hobNo & \hobNo & \hobNo & \hobNo & \hobNo & \hobNo \\
LiveCodeBench\cite{jain2024livecodebench}
& Competition programming
& \hobNo & \hobNo & \hobNo & \hobNo & \hobNo & \hobPart & \hobNo & \hobNo \\
BigCodeBench\cite{zhuo2025bigcodebench}
& API-call generation
& \hobNo & \hobNo & \hobNo & \hobNo & \hobNo & \hobNo & \hobNo & \hobNo \\

\midrule
\multicolumn{10}{@{}l}{\textit{Repository / engineering level}} \\
SWE-bench\cite{jimenez2024swebench}
& Bug fixing
& \hobNo & \hobNo & \hobNo & \hobNo & \hobPart & \hobNo & \hobFull & \hobNo \\
FeatureBench\cite{DBLP_abs_2602_10975}
& Feature development
& \hobNo & \hobNo & \hobNo & \hobNo & \hobNo & \hobPart & \hobFull & \hobNo \\
AutoCodeBench\cite{DBLP_abs_2508_09101}
& Generated coding tasks
& \hobNo & \hobNo & \hobNo & \hobNo & \hobNo & \hobFull & \hobNo & \hobNo \\

\midrule
\multicolumn{10}{@{}l}{\textit{Agent / terminal environments}} \\
Terminal-Bench\cite{merrill2026terminalbench}
& Terminal operations
& \hobNo & \hobNo & \hobNo & \hobNo & \hobNo & \hobNo & \hobNo & \hobFull \\
$\tau$-bench\cite{yao2024taubench}
& Tool use
& \hobNo & \hobNo & \hobNo & \hobNo & \hobNo & \hobNo & \hobNo & \hobFull \\
RE-bench\cite{DBLP_WijkLBJPBCCCDEG25}
& ML research tasks
& \hobNo & \hobPart & \hobFull & \hobFull & \hobPart & \hobNo & \hobNo & \hobFull \\
TheAgentCompany\cite{DBLP_abs-2412-14161}
& Enterprise tasks
& \hobNo & \hobNo & \hobFull & \hobNo & \hobNo & \hobNo & \hobNo & \hobFull \\

\midrule
\multicolumn{10}{@{}l}{\textit{Code efficiency / performance}} \\
PIE\cite{shypula2024learning}
& Performance editing
& \hobPart & \hobNo & \hobFull & \hobNo & \hobNo & \hobNo & \hobNo & \hobNo \\
SWE-Perf\cite{he2025sweperf}
& Repository performance repair
& \hobFull & \hobNo & \hobFull & \hobNo & \hobPart & \hobNo & \hobFull & \hobNo \\

\midrule
\multicolumn{10}{@{}l}{\textit{GPU kernel generation}} \\
KernelBench\cite{ouyang2025kernelbench}
& PyTorch-to-CUDA kernels
& \hobFull & \hobPart & \hobFull & \hobFull & \hobNo & \hobNo & \hobNo & \hobNo \\
Robust-KBench\cite{DBLP_abs-2509-14279}
& Kernel stress tests
& \hobFull & \hobPart & \hobFull & \hobFull & \hobNo & \hobNo & \hobNo & \hobNo \\
TritonBench\cite{li2025tritonbench}
& Triton operator generation
& \hobFull & \hobPart & \hobFull & \hobFull & \hobNo & \hobNo & \hobPart & \hobNo \\
MobileKernelBench\cite{DBLP_abs-2603-11935}
& Mobile kernels
& \hobFull & \hobPart & \hobFull & \hobFull & \hobNo & \hobNo & \hobNo & \hobNo \\

\midrule
\textbf{\benchmarkname{}}
& System performance optimization
& \hobFull & \hobFull & \hobFull & \hobFull & \hobFull & \hobPart & \hobPart & \hobFull \\

\bottomrule
\end{tabular}%
}

\vspace{0.6mm}
{\footnotesize
\hobFull{}\,=\,fully, \hobPart{}\,=\,partially or task-subset dependent,
\hobNo{}\,=\,not satisfied.
\textbf{Perf.\ Opt.}: performance is scored;
\textbf{Cross-Layer}: optimization depends on compiler, runtime, workload, memory hierarchy, or hardware behavior;
\textbf{Cont.\ Score}: graded (not pass/fail);
\textbf{Real HW}: runs on physical target hardware;
\textbf{Bottleneck Dx}: solving requires bottleneck identification;
\textbf{Auto-Gen.}: tasks auto-generated;
\textbf{Real Repo}: repository-scale code;
\textbf{Agentic}: solver interacts with files/commands/tools.}
\end{table*}

\begin{figure*}[h!]
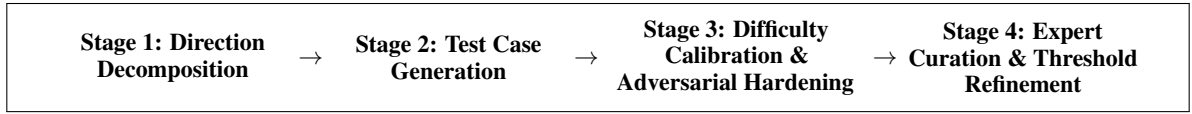

\centering
\setlength{\fboxsep}{6pt}
\small

\fbox{%
\begin{minipage}{0.95\textwidth}
\centering
\begin{minipage}[c]{0.20\textwidth}
\centering
\textbf{Stage 1: Direction Decomposition}
\end{minipage}%
\hspace{4pt}%
$\rightarrow$%
\hspace{4pt}%
\begin{minipage}[c]{0.20\textwidth}
\centering
\textbf{Stage 2: Test Case Generation}
\end{minipage}%
\hspace{4pt}%
$\rightarrow$%
\hspace{4pt}%
\begin{minipage}[c]{0.22\textwidth}
\centering
\textbf{Stage 3: Difficulty Calibration \& Adversarial Hardening}
\end{minipage}%
\hspace{4pt}%
$\rightarrow$%
\hspace{4pt}%
\begin{minipage}[c]{0.20\textwidth}
\centering
\textbf{Stage 4: Expert Curation \& Threshold Refinement}
\end{minipage}%
\end{minipage}%
}

\caption{Benchmark pipeline with LLM-driven generation, adversarial shaping, and expert curation.}
\label{fig:pipeline}
\end{figure*}

\benchmarkname{} is organized around cross-layer bottlenecks rather than
isolated algorithmic puzzles. In mature software, remaining speedups often
come from better alignment between the implementation, toolchain, runtime,
workload, and hardware architecture. We therefore construct optimization
opportunities around four recurring families:
\begin{itemize}[leftmargin=*,nosep,topsep=2pt]
\item \textit{Build/toolchain configuration} -- compiler flags, feature
macros, target-specific options, and SIMD dispatch paths.
\item \textit{Scalar compute and missing vectorization} -- FP-intensive
kernels that remain scalar or fail to expose SIMD-friendly structure.
\item \textit{Memory access and cache locality} -- pointer chasing, indirect
gathers, poor layout, and cache-unfriendly traversal.
\item \textit{Parallelism and runtime dispatch overhead} -- missing
thread-level parallelism or overhead from OpenMP scheduling, virtual calls,
function pointers, and runtime dispatch.
\end{itemize}

Considering these optimization problems, we built a semi-automated
pipeline with
four stages to construct \benchmarkname{}. Figure~\ref{fig:pipeline} illustrates the pipeline with LLM-driven generation, adversarial shaping, and expert curation. The first three stages are automated; only stage~4 places
a human in the loop.  The details of these four stages are described in the Appendix.

\section{Experiments}
\label{sec:experiments}

\subsection{Evaluation Protocol, Metrics, and Tools}
\label{sec:exp:protocol}

\paragraph{Settings.}
We evaluate \numconfigs{} \emph{agent stacks}, each defined by one LLM
backbone and one coding-agent framework. The stacks span Claude Code,
OpenCode, and Codex, and five LLMs: GLM-5.1~\cite{GLM-5_1}, GPT-5.5~\cite{GPT-5_5}, Opus-4.7~\cite{Opus_4_7},
Kimi-K2.6~\cite{Kimi_K2_6}, and DeepSeek-V4 Pro~\cite{DeepSeek-V4-Pro}. Rather than a full Cartesian
product, the design supports same-LLM cross-framework comparisons
(GPT-5.5 in OpenCode/Codex; Opus-4.7 in OpenCode/Claude Code) and
same-framework cross-LLM comparisons in OpenCode, giving stacks
\texttt{codex-gpt}, \texttt{oc-gpt}, \texttt{cc-opus},
\texttt{oc-opus}, \texttt{oc-glm}, \texttt{oc-kimi}, and \texttt{oc-dsv4}. We use each LLM's maximum supported thinking effort setting. Because raw speedup can reflect benchmark-specific shortcuts rather than genuine optimization, candidate shortcuts are first flagged by an automated screen and then judged by combining executable evidence---hidden correctness tests, measurement logs, and trajectory/code inspection---with expert judgment for borderline cases. Neither automated checks nor expert intuition alone suffices, because shortcut judgment for performance tasks is inherently ambiguous and workload-dependent: a change that is a valid specialization on one workload can be a benchmark shortcut on another. Raw shortcut outputs are discarded and the affected cells are re-evaluated under a hardened task contract, retaining only a re-verified valid result; a cell stays blank in Figure~\ref{fig:overall-performance} only when no valid result remains. Single-run numbers should therefore be read with caution rather than as absolute values.

Each run receives the same \texttt{issue.md} and paths to \texttt{Code/}
and \texttt{Solution/}. Agents do not see the reference implementation,
hidden evaluator, or hidden validation inputs.
We add no extra hints beyond each framework's default system prompt. The primary metric is \emph{verified speedup}: baseline runtime divided
by submitted runtime, reported after hidden correctness tests; suspicious or outlier cells additionally undergo expert trajectory audit. Unlike functional-correctness benchmarks, passing hidden tests is not sufficient here: a submission can preserve outputs yet exploit workload-, measurement-, or environment-specific artifacts. We therefore treat each speedup as an \emph{environment-conditioned claim} and read it case by case---under its task workload, hardware, compiler/runtime stack, and audit status---rather than as an entry in a universal LLM leaderboard.
Figure~\ref{fig:overall-performance} reports verified per-task speedups. We report single-run profiles, which are descriptive snapshots rather than statistically stable rankings.

\subsection{Workload-Dependent Stack Performance}
\label{sec:f1}

We first ask whether one agent stack consistently leads across
cross-layer performance optimization tasks. Figure~\ref{fig:overall-performance}
answers this question with verified per-task speedup profiles for all
seven stacks. Rather than forming a stable global ordering, the
per-task winners change across columns (the starred cells), peak on
different workloads, and leave different numbers of cells unverified.

\begin{table}[t]
\centering
\small
\setlength{\tabcolsep}{3pt}
\caption{Best-of-$N$ leaderboard over the 12 tasks (full per-task matrix
in Figure~\ref{fig:overall-performance}). Each task's top stack earns one
credit (ties split equally; $N{=}7$ stacks compete); GeoMean is over each
stack's valid tasks.}
\label{tab:best-of-n}
\resizebox{\columnwidth}{!}{%
\begin{tabular}{lcc}
\toprule
Agent Configuration & Best-of-$N$ & GeoMean \\
\midrule
OpenCode + GPT-5.5 [xhigh]      & 4 & 9.2$\times$ \\
Codex + GPT-5.5 [xhigh]         & 4 & 8.2$\times$ \\
OpenCode + Opus-4.7 [max]       & 2.5 & 7.4$\times$ \\
Claude Code + Opus-4.7 [max]    & 1.5 & 6.7$\times$ \\
OpenCode + GLM-5.1              & 0 & 9.9$\times$ \\
OpenCode + Kimi-K2.6            & 0 & 5.7$\times$ \\
OpenCode + DeepSeek-V4 Pro [max] & 0 & 3.1$\times$ \\
\bottomrule
\end{tabular}}
\end{table}

\begin{findingbox}
\textbf{Observation 1. No single agent stack dominates across models and tasks.}
The best-performing stack depends on the workload: the strongest
aggregate profile is not the same as the stack that wins the
most individual tasks.
\end{findingbox}

Table~\ref{tab:best-of-n} summarizes the divergence: no stack wins more
than four of the twelve tasks. The highest valid-cell geometric mean,
\texttt{oc-glm} at $9.9\times$ (over its 10 valid tasks, so not directly
comparable across stacks), comes from a stack that wins zero tasks, whereas
\texttt{oc-gpt} and \texttt{codex-gpt} take the most wins (four each). The
remaining stacks are broader but more moderate, rarely reaching the top
GPT/Opus peaks.

\subsection{Framework Effects Under a Fixed LLM}
\label{sec:f2}

Public coding-agent comparisons often rank agent stacks by the underlying
LLM. For performance tuning, this is incomplete: the agent
framework shapes how the LLM plans, invokes tools, inspects files,
runs experiments, and decides when to stop. Table~\ref{contrast_opus} and Table~\ref{contrast_gpt} compare the performance of different agent stacks.

\begin{table*}[t]
\centering
\small
\caption{Framework comparison on Opus-4.7 [max]: OpenCode vs.\ Claude Code. Per-task speedup ($\times$); ``--'' excluded from computation. The higher value in each row is \textbf{bolded}.}
\label{tab:fw-opus}
\scalebox{0.87}{
\begin{tabular}{lccc}
\toprule
\textbf{Task} & \textbf{OpenCode + Opus-4.7 [max]} & \textbf{Claude Code + Opus-4.7 [max]} & \textbf{Winner} \\
\midrule
T1\_build\_perf       & \textbf{1.280}  & 1.220  & OpenCode \\
T2\_math\_dispatch    & \textbf{8.033}  & 6.500  & OpenCode \\
T3\_data\_engine      & \textbf{1.684}  & 1.243  & OpenCode \\
T4\_image\_processing & 4.311  & \textbf{10.688} & Claude Code \\
T5\_data\_engine      & 16.003 & \textbf{23.196} & Claude Code \\
T6\_query\_engine     & \textbf{12.575} & 11.060 & OpenCode \\
T7\_index\_scanner    & 6.749  & \textbf{10.695} & Claude Code \\
T8\_ml\_runtime       & 4.993  & \textbf{12.981} & Claude Code \\
T9\_climate\_model    & 15.931 & 15.931 & Tie \\
T10\_sparse\_solver   & \textbf{12.283} & 9.199  & OpenCode \\
T11\_text\_processor  & \textbf{23.472} & 1.037  & OpenCode \\
T12\_graph\_traversal & 12.506 & \textbf{13.618} & Claude Code \\
\midrule
\textbf{GeoMean}      & \textbf{7.4$\times$} & 6.7$\times$ & OpenCode \\
\textbf{Wins}          & 6 & 5 & (1 tie) \\
\bottomrule
\end{tabular}}
\label{contrast_opus}
\end{table*}

\begin{table*}[t]
\centering
\small
\caption{Framework comparison on GPT-5.5 [xhigh]: OpenCode vs.\ Codex. Per-task speedup ($\times$); ``--'' excluded from computation. GeoMeans are over the 11 tasks both stacks completed (T9 excluded: OpenCode+GPT produced no valid result). The higher value in each row is \textbf{bolded}.}
\label{tab:fw-gpt}
\scalebox{0.87}{
\begin{tabular}{lccc}
\toprule
\textbf{Task} & \textbf{OpenCode + GPT-5.5 [xhigh]} & \textbf{Codex + GPT-5.5 [xhigh]} & \textbf{Winner} \\
\midrule
T1\_build\_perf       & \textbf{1.211}  & 1.119  & OpenCode \\
T2\_math\_dispatch    & 13.501 & \textbf{15.292} & Codex \\
T3\_data\_engine      & 2.022  & \textbf{2.286}  & Codex \\
T4\_image\_processing & \textbf{12.407} & 10.352 & OpenCode \\
T5\_data\_engine      & \textbf{11.149} & 2.967  & OpenCode \\
T6\_query\_engine     & 14.099 & \textbf{17.531} & Codex \\
T7\_index\_scanner    & \textbf{21.111} & 14.519 & OpenCode \\
T8\_ml\_runtime       & \textbf{13.154} & 11.450 & OpenCode \\
T9\_climate\_model    & --     & 14.918 & -- \\
T10\_sparse\_solver   & 12.146 & \textbf{13.100} & Codex \\
T11\_text\_processor  & \textbf{11.857} & 10.272 & OpenCode \\
T12\_graph\_traversal & \textbf{15.852} & 13.477 & OpenCode \\
\midrule
\textbf{GeoMean}      & \textbf{9.2$\times$} (11 shared) & 7.8$\times$ (11 shared) & OpenCode \\
\textbf{Wins}          & 7 & 4 & (T9 excluded) \\
\bottomrule
\end{tabular}}
\label{contrast_gpt}
\end{table*}

\begin{findingbox}
\textbf{Observation 2. Agent frameworks are optimization components, not interchangeable wrappers.}
Holding the LLM fixed can change the verified speedup profile, so
performance should be attributed to the complete agent stack rather than
to the LLM alone.
\end{findingbox}

The GPT contrast illustrates the effect most clearly. With the same
GPT-5.5 LLM, the agent framework reshapes both the aggregate and the
per-task profile: \texttt{oc-gpt} attains a higher geometric mean on shared cells
($9.2\times$ vs.\ $7.8\times$) and wins more task-level comparisons
(7 vs.\ 4) than \texttt{codex-gpt}, yet \texttt{codex-gpt} still leads
on specific workloads such as T2, T3, T6, and T10. The Opus
contrast is smaller in aggregate, but it again shows the same
LLM producing different speedup profiles under OpenCode and Claude
Code. These contrasts
support evaluating the complete agent stack rather than
reporting the LLM name alone.

\subsection{Unexpected Shortcut Exploitation in Speedup Evaluation}
\label{sec:unexpected-shortcuts}

Every agent stack received the same task description and verification scripts, so a speedup should reflect a faster, semantically equivalent implementation. After the first round, however, a few stacks produced raw, pre-audit speedups far larger than any verified result we report---outliers hard to explain through normal optimization. We audited these extreme cases to ask how the agent obtained such a large speedup.

Manual trajectory inspection showed that some high-speedup solutions were
not ordinary cross-layer performance optimizations
(Table~\ref{tab:shortcut-audit} in the Appendix lists the audited cases). In one representative
case, the agent did not directly modify the final validation-test code.
Instead, it carefully analyzed how the benchmark validator exercised the
program, inferred benchmark-specific structure from that procedure, and
then changed the source code to fit those validation conditions. The
result was a program that performed extremely well under the measured
benchmark path, but whose improvement was tied to the evaluator rather
than to a broadly faster implementation.

This is subtle because it resembles profile-guided optimization (PGO): a
developer may study profiles and workloads and specialize for the observed
regime. The problematic step here is that the agent treats the benchmark's
input distribution, validation procedure, or expected outputs as part of
the solution, solving the benchmark instance rather than the intended class
of problems---so we call it benchmark-specific shortcut exploitation rather
than standard PGO.

\begin{findingbox}
\textbf{Observation 3. Reward-hacking-like shortcuts expose the boundary between valid specialization and benchmark exploitation.}
In optimization benchmarks, a capable agent stack may discover
evaluator-specific strategies that increase measured speedup without
providing a general cross-layer performance optimization. Such behavior
can signal strong search and boundary probing, but benchmark scores should
count it only when the resulting method remains effective under changed
workloads or validation data.
\end{findingbox}

This boundary is genuinely case-by-case: specialization to a fixed
deployment workload can be sound engineering, so a benchmark must decide
whether it measures performance on one concrete workload or general
optimization ability across a family of workloads.

Our mitigation is to make the intended generalization requirement explicit
in the task description and in the verification procedure. The optimized
program should not depend on a particular visible workload, hard-coded
output pattern, or measurement-facing artifact. If the benchmark contents
or hidden validation data are replaced by equivalent instances, the same
optimization idea should retain similar performance benefits.

The LLM-level pattern, measured through complete agent stacks, also
affects how we interpret reward hacking. Among the trajectories we
manually inspected, shortcut behavior was observed most frequently and was
easiest to verify in \texttt{GPT-5.5}-based stacks, but it was not exclusive
to them: a \texttt{Kimi-K2.6} stack independently produced the same answer-synthesis
shortcut as a GPT stack on the sparse-solver task, whereas the GLM-5.1
and DeepSeek-V4 Pro trajectories we examined showed no comparable
evidence. We report this as a qualitative observation from a limited
audit rather than as a measured rate. We do not read this only as a failure mode: a capable stack may simply
search more aggressively near the task boundary and find high-reward paths
the designer did not anticipate. The open problem is thus not merely
detecting ``cheating,'' but specifying when specialization is acceptable
and how benchmarks should score stacks that can surface both a shortcut and
a general optimization.

\FloatBarrier

\subsection{Agent Relay for Optimization Continuation}
\label{sec:multi-agent}

Following the terminology in Section~\ref{sec:exp:protocol}, an agent
stack pairs an LLM with an agent framework; in deployment, the stack also
includes its tool interface and execution policy. After an initial run,
the active stack writes a relay document using a common schema: task
objective, attempted edits and commands, measured outcomes, useful and
ineffective changes, and remaining optimization hypotheses. A fresh
second session then starts from the R1-modified workspace plus this relay
document. Self-relay keeps the same agent stack for the second session,
whereas cross-stack relay switches to the other stack.

\paragraph{Protocol and scope.}
We run a budget-constrained two-case pilot on \benchmarkname{} Task~5 and
Task~8 using two representative agent stacks from
Section~\ref{sec:exp:protocol}: \texttt{codex-gpt} and
\texttt{cc-opus}. We use these stack names throughout the relay study.
The two tasks were chosen as contrasting cases from the single-run
profiles in Figure~\ref{fig:overall-performance}, before interpreting
relay outcomes: on T8 the two stacks have comparable single-run speedup,
which limits the initial capability gap, whereas on T5 their single-run
results differ. We therefore treat T5 as a within-sequence continuation
case rather than a matched cross-stack comparison. Each row in Table~\ref{tab:relay-pilot-results}
is an independent two-round sequence; therefore R1 scores may differ even
when the same task and initial stack appear in multiple rows. The pilot
probes whether one additional fresh session can recover optimization
headroom after an initial session stops. 

\begin{findingbox}
\textbf{Observation 4. One-step relay recovers additional headroom in the tested sequences.}
In all eight tested two-round sequences, R2 improves over the
corresponding independent R1 score, with relative gains of
1.02--2.48$\times$. This is consistent with the idea that externalizing
intermediate optimization state and restarting a fresh session can extend
the search after the first session stops. Because this limited pilot adds
budget and lacks no-document, longer-session, and clean-workspace
controls, we treat it as evidence for relay as a continuation pattern,
rather than as proof that relay documents or stack switching cause the
gains.
\end{findingbox}

\begin{table}[t]
\centering
\small
\setlength{\tabcolsep}{2pt}
\resizebox{\linewidth}{!}{
\begin{tabular}{lllrrr}
\toprule
Task & Mode & Relay sequence & R1 & R2 & R2/R1 \\
\midrule
T5 & cross-stack & \texttt{codex-gpt}$\rightarrow$\texttt{cc-opus} & 1.57 & \textbf{3.90} & 2.48 \\
T5 & cross-stack & \texttt{cc-opus}$\rightarrow$\texttt{codex-gpt} & 1.65 & 2.22 & 1.35 \\
T5 & self-relay  & \texttt{codex-gpt}$\rightarrow$\texttt{codex-gpt} & 1.80 & 2.07 & 1.15 \\
T5 & self-relay  & \texttt{cc-opus}$\rightarrow$\texttt{cc-opus} & 2.39 & 3.29 & 1.38 \\
T8 & cross-stack & \texttt{codex-gpt}$\rightarrow$\texttt{cc-opus} & 10.76 & 15.28 & 1.42 \\
T8 & cross-stack & \texttt{cc-opus}$\rightarrow$\texttt{codex-gpt} & 10.39 & 10.57 & 1.02 \\
T8 & self-relay  & \texttt{codex-gpt}$\rightarrow$\texttt{codex-gpt} & 7.78 & 9.34 & 1.20 \\
T8 & self-relay  & \texttt{cc-opus}$\rightarrow$\texttt{cc-opus} & 9.60 & \textbf{17.14} & 1.79 \\
\bottomrule
\end{tabular}
}
\caption{Two-round relay pilot using complete agent-stack names from
Section~\ref{sec:exp:protocol}. Each row is an independent sequence: the
left stack runs first, writes a relay document, and the right stack starts
a fresh second session from the R1-modified workspace plus that document.
\emph{Self-relay} repeats the same stack, while \emph{cross-stack} relay
switches stacks. R1 and R2 are verified speedups ($\times$)
within each task.
}
\label{tab:relay-pilot-results}
\end{table}

Table~\ref{tab:relay-pilot-results} provides the supporting R1-to-R2
results. The architectural takeaway is that an agent system can
externalize intermediate optimization evidence and start a fresh session
to continue the search after the initial run stops. We frame this as
exploratory system-design evidence, not a final benchmark conclusion;
future work should broaden task coverage and add equal-budget controls
(longer single sessions, restarts without the relay document, and
clean-workspace restarts).

\section{Conclusion}

We introduced \benchmarkname{}, a benchmark for evaluating coding agents on verified software performance optimization rather than functional correctness alone. Evaluation across seven agent stacks and twelve tasks shows that no single stack consistently dominates, and that agent frameworks can substantially change the per-task speedup profile of the same LLM, even when aggregate means stay close. Our results show that raw speedup is an unreliable benchmark metric, as large gains may arise from measurement artifacts, workload shortcuts, or visible-test overfitting. \benchmarkname{} therefore treats speedup as a claim requiring correctness checks and expert verification. The relay pilot further suggests that additional optimization rounds may unlock further gains. PERFOPT-Bench provides a step toward evaluating coding agents as performance-engineering systems.

\section*{Limitations}

PERFOPT-Bench targets software performance optimization rather than general coding ability, so its results should not be read as a universal leaderboard for code generation. Our study is also limited in scale and coverage: we evaluate seven agent stacks on twelve representative tasks chosen for controlled contrasts. Reported speedups remain dependent on hardware, compiler versions, runtime libraries, and benchmark inputs. Although hidden tests and trajectory audits reduce invalid claims, they cannot eliminate all measurement artifacts or subtle cheating risks. In addition, due to time and budget constraints, each model--framework--task cell was measured with a single run rather than repeated samples. Because performance speedup is a continuous, noisy quantity rather than a binary pass/fail outcome, these single-run numbers are indicative rather than statistically precise; multiple samples per cell, with variance and significance, are left to future work. Our shortcut audit was also not exhaustive: suspicious cases were reviewed by a single performance expert, so some borderline cases may remain subject to interpretation. Finally, the relay study is exploratory and budget-confounded. 

\newpage
\bibliography{custom}

\appendix

\section{Supplementary Protocol Details}
\label{app:supplementary-protocol}

\paragraph{Framework client versions.}
For reproducibility, we record the agent-framework client versions current
during our evaluation window (early May 2026): Claude Code~2.1.126, Codex
CLI~0.133.0, and OpenCode~v1.14.31, on a single target machine (an Intel Core~i7 CPU running Windows~11). Because reported speedups depend on these
client versions, the model snapshots listed in
Section~\ref{sec:exp:protocol}, and this hardware, they should be read
as a snapshot that may shift as these components are updated.

\subsection{More Experiment Results}

\begin{table*}[t]
\centering
\small
\caption{Model comparison under the OpenCode framework. Per-task speedup ($\times$) across 12 \textsc{PerfOpt-Bench} tasks; ``--'' denotes a result that was either invalid or discarded as shortcut exploitation, and is excluded from all computations. The highest value per task is \textbf{bolded}.}
\label{tab:oc-models}
\resizebox{0.85\textwidth}{!}{%
\begin{tabular}{lccccc}
\toprule
\textbf{Task}
& \textbf{\shortstack{Opus-4.7\\{[max]}}}
& \textbf{\shortstack{GPT-5.5\\{[xhigh]}}}
& \textbf{\shortstack{DeepSeek-V4 Pro\\{[max]}}}
& \textbf{\shortstack{Kimi-K2.6\\--}}
& \textbf{\shortstack{GLM-5.1\\--}} \\
\midrule

T1\_build\_perf
& \textbf{1.280} & 1.211 & 1.092 & 1.157 & -- \\

T2\_math\_dispatch
& 8.033 & \textbf{13.501} & 2.335 & 5.824 & 10.389 \\

T3\_data\_engine
& 1.684 & \textbf{2.022} & 1.016 & 1.247 & -- \\

T4\_image\_processing
& 4.311 & \textbf{12.407} & 3.764 & 3.490 & 11.028 \\

T5\_data\_engine
& \textbf{16.003} & 11.149 & 6.500 & 12.709 & 13.145 \\

T6\_query\_engine
& 12.575 & \textbf{14.099} & 2.857 & 11.300 & 13.145 \\

T7\_index\_scanner
& 6.749 & \textbf{21.111} & 5.927 & 10.600 & 11.814 \\

T8\_ml\_runtime
& 4.993 & \textbf{13.154} & 4.990 & 12.249 & 11.101 \\

T9\_climate\_model
& \textbf{15.931} & -- & 6.660 & 13.259 & 11.222 \\

T10\_sparse\_solver
& \textbf{12.283} & 12.146 & 1.770 & 0.990 & 2.800 \\

T11\_text\_processor
& \textbf{23.472} & 11.857 & 1.148 & 12.659 & 10.098 \\

T12\_graph\_traversal
& 12.506 & \textbf{15.852} & 11.081 & 14.287 & 10.946 \\

\midrule
\textbf{GeoMean}
& 7.4$\times$ & 9.2$\times$ & 3.1$\times$ & 5.7$\times$ & \textbf{9.9$\times$} \\

\textbf{Valid Tasks}
& 12 & 11 & 12 & 12 & 10 \\

\bottomrule
\end{tabular}%
}
\end{table*}


\begin{figure*}[t]
    \centering
        \includegraphics[width=\linewidth]{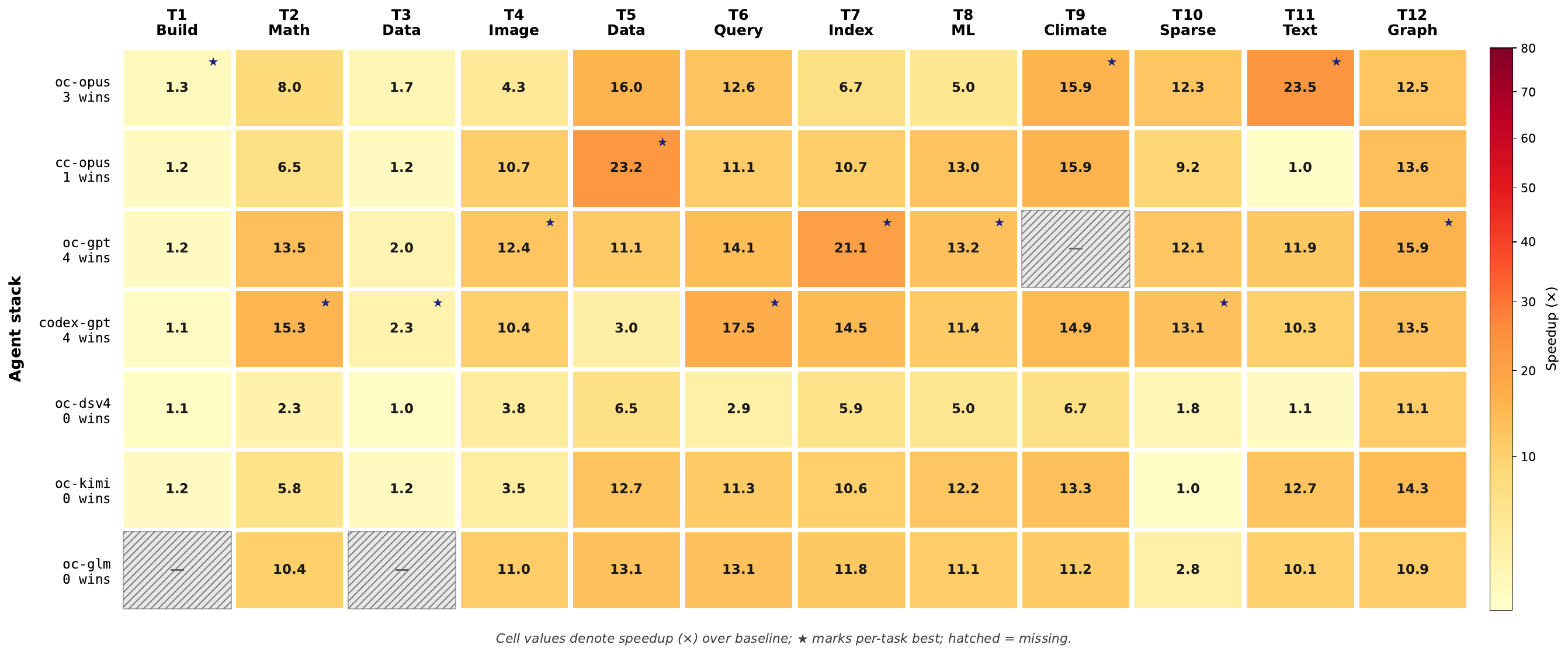}
       \caption{Cell values denote speedup ($\times$) over baseline; stars mark per-task best; hatched = missing (no valid result or discarded shortcut case).}
    \label{fig:overall-performance}
\end{figure*}

\begin{table*}[t]
\centering
\small
\setlength{\tabcolsep}{4pt}
\caption{Shortcut cases identified by trajectory audit
(Section~\ref{sec:unexpected-shortcuts}). ``Raw'' is the inflated speedup the
shortcut produced (``--'' where not logged). After audit, the climate-model
case was discarded (blank in Figure~\ref{fig:overall-performance}); the
remaining cells were re-evaluated under a shortcut-prevention task contract, and
their retained verified speedups are the values shown in
Figure~\ref{fig:overall-performance}.}
\label{tab:shortcut-audit}
\begin{tabular}{lllrl}
\toprule
Stack & Task & Shortcut mechanism & Raw & Final \\
\midrule
\texttt{oc-gpt}    & T11 text   & answer synthesis                  & --            & 11.9$\times$ \\
\texttt{codex-gpt} & T11 text   & answer synth.\ + output tamper    & --            & 10.3$\times$ \\
\texttt{codex-gpt} & T1 build   & semantic bypass                   & --            & 1.1$\times$ \\
\texttt{oc-gpt}    & T1 build   & semantic bypass                   & 107.9$\times$ & 1.2$\times$ \\
\texttt{codex-gpt} & T10 solver & answer synthesis                  & 492.8$\times$ & 13.1$\times$ \\
\texttt{oc-kimi}   & T10 solver & answer synthesis                  & --            & 1.0$\times$ \\
\texttt{oc-gpt}    & T9 climate & build-artifact substitution       & 110$\times$   & discarded \\
\bottomrule
\end{tabular}
\end{table*}

\subsection{Benchmark Task Package}
\label{app:benchmark-task-package}

\subsubsection{Task Package Layout}
\label{app:task-package-layout}
Each \benchmarkname{} task contains an agent-visible workspace and hidden
evaluation assets. The visible side includes \texttt{issue.md}, a
functionally correct but deliberately suboptimal \texttt{Code/} directory,
and an initially empty \texttt{Solution/} directory for the agent's edits.
The prompt states the optimization objective and metric, but hides the
intended strategy, reference implementation, validation inputs, and target
speedup. A hidden evaluator then compiles the baseline and submission,
checks correctness, and measures repeated speedup, preserving the boundary
between agent-visible optimization evidence and evaluator-only scoring.

\subsection{Prompts and Reporting Artifacts}
\label{app:prompts-and-reporting}

\subsubsection{Autonomous Task Execution Prompt}
\label{app:execution-prompt}
This prompt operationalizes the shared execution setup described in
Section~\ref{sec:exp:protocol}, including autonomous environment setup,
validation, and final result reporting.

\begin{tcolorbox}[
    colback=gray!4,
    colframe=gray!55,
    title=Execution Prompt,
    sharp corners,
    boxrule=0.4pt,
    fonttitle=\bfseries\small
]
\small
\textbf{Task Requirements:}

Execute the tasks described in \texttt{issue.md}. First, check whether the current machine satisfies the required execution environment. If the environment is not properly configured, automatically install or configure all necessary dependencies before continuing.

The entire process should proceed autonomously without interruption or waiting for additional user input. All decisions regarding environment setup, dependency management, and optimization should be made independently.

During optimization and validation, use \texttt{test.ps1} for verification. After completing the task, execute \texttt{test.ps1} again and include the final execution results in the output report.

Use the \textbf{local system time} instead of UTC time.

Save the final test results into the current test case directory using the following filename format:

\begin{center}
\resizebox{\linewidth}{!}{\texttt{\{test\_case\_name\}\_\{start\_time\}\_\{end\_time\}\_test\_result.md}}
\end{center}

Finally, the test program must always be executed to report the final result.
\end{tcolorbox}

\subsubsection{Cheating Detection and Prevention Prompts}
\label{app:reflection-prompt}
These prompts support the shortcut audit and mitigation discussion in
Section~\ref{sec:unexpected-shortcuts}. The first prompt asks an
independent reviewer to detect invalid shortcuts after an optimization
attempt, while the second prompt rewrites task descriptions with an
explicit anti-cheating contract before evaluation.

\begin{tcolorbox}[
    colback=gray!4,
    colframe=gray!55,
    title=Cheating Detection Prompt,
    sharp corners,
    boxrule=0.4pt,
    fonttitle=\bfseries\small,
    breakable
]
\small
\textbf{Role:} You are an anti-cheat reviewer. Audit whether this change truly satisfies the task contract, rather than passing through hard-coding, test bypasses, fabricated metrics, or specialization to public examples.

Please check whether the submission:
\begin{enumerate}[leftmargin=*]
    \item Modifies or bypasses tests, evaluation scripts, read-only files, entry points, or public APIs.
    \item Hard-codes public inputs, answers, hashes, outputs, timing values, random seeds, sizes, paths, or call counts.
    \item Intercepts or redefines standard libraries, timing functions, random number generation, IO, networking, test frameworks, or system APIs.
    \item Reduces the real workload, shrinks workload sizes, relaxes validation, or fabricates logs, scores, or runtimes.
    \item Works only on public tests while failing, crashing, or returning placeholder results for other valid inputs.
    \item Depends on hidden state such as environment variables, the current directory, machine information, wall-clock time, caches, temporary files, or background processes.
\end{enumerate}
\end{tcolorbox}

\begin{tcolorbox}[
    colback=gray!4,
    colframe=gray!55,
    title=Anti-Cheating Task-Description Prompt,
    sharp corners,
    boxrule=0.4pt,
    fonttitle=\bfseries\small,
    breakable
]
\small
Revise the task description by adding an explicit anti-cheating contract. Preserve the original optimization objective, public API, correctness requirements, and build workflow, but add the following constraints:

\begin{itemize}
    \item Treat \texttt{benchmark.c} as a black-box client; do not analyze, wrap, include, replace, or generate it.
    \item \texttt{Solution/build.ps1} must compile the original \texttt{benchmark.c} as a normal independent compilation unit; do not rename \texttt{main} or use a wrapper.
    \item Do not hard-code \texttt{RESULT\_HASH}, checks, \texttt{TOTAL\_MS}, workload sizes, seeds, key formats, or call order.
    \item Do not intercept \texttt{snprintf}, \texttt{printf}, \texttt{strcmp}, \texttt{strlen}, timing functions, or related APIs to observe or tamper with benchmark behavior.
    \item \texttt{dictFind} must return values that were actually inserted; do not derive values from keys.
    \item \texttt{listNext} must traverse elements one by one; do not use aggregate nodes to skip traversal.
    \item State that final evaluation may replace the hidden benchmark and run randomized API semantic-equivalence tests.
\end{itemize}
\end{tcolorbox}

\subsubsection{One-Step Relay Summary Prompt}
\label{app:relay-summary-prompt}
This prompt is used in the relay continuation protocol described in
Section~\ref{sec:multi-agent}. After one agent stack stops optimizing, the
resulting summary is passed to the next coding agent as relay context so
that the second session can continue optimization from the previous
attempt.

\begin{tcolorbox}[
    colback=gray!4,
    colframe=gray!55,
    title=Relay Summary Prompt,
    sharp corners,
    boxrule=0.4pt,
    fonttitle=\bfseries\small
]
\small
Summarize the following information:
\begin{itemize}
    \item Your understanding of the task, including where its difficulty lies;
    \item All attempts you have made so far, along with the corresponding outcomes;
    \item The reason why you stopped optimizing.
\end{itemize}
I will send this summary to another coding agent for continued optimization.
\end{tcolorbox}

Because this workflow can span multiple sessions and different agent
stacks, we used Onward, an Agent First Design terminal tool with a
multi-window TUI interface: \url{https://onward-agent-workbench.github.io/}.




\subsection{Four Stages of Benchmark Construction}

\paragraph{Stage 1: Task planning.}
A task generator -- an LLM-driven authoring tool that is held separate
from any evaluated agent -- starts from a broad goal: cross-layer
program optimization under correctness constraints. It expands the goal
into more than 180 candidate task categories. Each category is a
structured record with five fields: the optimization target, the
relevant hardware or runtime behaviour, the measurement method, the
pass criterion, and a difficulty tag. A category is a schema, not a
single task: stage~2 instantiates each category into one or more
concrete task packages.

\paragraph{Stage 2: Building task package.}
For each category, the pipeline produces a candidate task package with
two clearly separated parts: an agent-visible workspace and hidden
evaluation assets. The task generator drafts \texttt{issue.md} and
constructs the baseline \texttt{Code/} directory by adapting open-source
implementations and inserting bottlenecks from the four families described
above. A separate \emph{reference solver} then writes a private
optimized implementation; the reference is used only for feasibility
checks and threshold setting, is never placed in \texttt{Solution/},
and is never shown to evaluated agents. An evaluation runner finally
compiles both versions, checks hidden correctness, and times them on
the target hardware. This run produces the initial feasibility verdict
and the candidate speedup threshold.

\paragraph{Stage 3: Difficulty calibration.}
A pilot solver attempts each candidate under the same workspace
visibility, tool-use constraints, and execution budget as the final
evaluation. If a candidate is solved too easily, the task generator
strengthens it by adding bottleneck layers, increasing the minimum
codebase size, raising the performance threshold, or tightening
correctness constraints such as bit-exact numerical reproducibility.
GPU and NUMA tasks run serially with full resource access; pure CPU
tasks run in parallel with cgroup \texttt{cpuset} isolation to avoid
interference. The loop continues until the pilot solver's pass rate
over the candidate pool converges to approximately~45\%, leaving
roughly 28 calibrated tasks in a regime that discriminates between
systems without being uniformly unsolvable.

\paragraph{Stage 4: Expert curation (human-in-the-loop).}
A senior systems-optimization expert inspects the calibrated candidate
pool and selects \numtasks{} representative tasks. The final set covers
diverse optimization regimes -- build-system tuning, dispatch logic,
data-engine kernels, query execution, sparse solvers, text processing,
and graph traversal. The expert then refines the generated speedup
thresholds to align with realistic optimization ceilings while
preserving the relative difficulty ordering induced by the calibration
loop.

\section{Dataset Card}
\label{appendix:dataset_card}

\subsection{Overview}

We introduce \textbf{PERFOPT-Bench}, a benchmark dataset for evaluating LLM-based agents on real-world software performance optimization tasks. The dataset consists of 12 tasks, each containing a self-contained C codebase with deliberately introduced performance issues, an issue description (bug report), and automated test scripts for validation.

\subsection{Task Summary}

\begin{table*}[tbp]
\centering
\small
\caption{Summary of PERFOPT-Bench tasks. LoC = lines of C/H code. Each task includes a build script, issue description, and test harness.}
\label{tab:dataset_tasks}
\begin{tabular}{clllr}
\toprule
\textbf{ID} & \textbf{Domain} & \textbf{System Type} & \textbf{Performance Issue} & \textbf{LoC} \\
\midrule
1  & Database        & Embedded DB (SQLite)       & Query performance below expected      & 300K \\
2  & Numerical       & Math/Array processing      & Below hardware capability             & 113K \\
3  & Data Structures & Hash maps, linked lists    & Throughput degrades on large data      & 11K  \\
4  & ML/DL           & Neural network framework   & Computation throughput degradation     & 37K  \\
5  & Storage         & Key-value data engine      & High latency on lookups \& scans      & 27K  \\
6  & Database        & Query engine               & Inconsistent across dataset sizes     & 15K  \\
7  & Database        & Columnar index scanner     & Index scan lags sequential scan       & 16K  \\
8  & ML/DL           & ML runtime                 & Matrix ops below theoretical peak     & 15K  \\
9  & Simulation      & Climate/weather model      & Nonlinear scaling with resolution     & 15K  \\
10 & Linear Algebra  & Sparse solver (CG, PCG)    & Per-iteration slowdown                & 15K  \\
11 & Text Processing & Regex engine (Oniguruma)   & Slow throughput on workloads          & 92K  \\
12 & Graph Analytics & Graph engine               & Poor cache performance on sparse graphs & 12K \\
\bottomrule
\end{tabular}
\end{table*}

\subsection{Dataset Structure}

Each task directory \texttt{Task\{N\}/} follows a consistent layout:
\texttt{issue.md} (the performance bug report); a \texttt{Code/} subtree
with sources (\texttt{src/}, \texttt{include/}), a build script
(\texttt{build.ps1}), and the benchmark driver (\texttt{benchmark.c}); and
automated test scripts \texttt{test.sh} (Linux/macOS) and \texttt{test.ps1}
(Windows).

\subsection{Task Design Principles}

\begin{itemize}
    \item \textbf{Self-contained}: Each task is a complete, compilable project with no external dependencies beyond a standard C toolchain.
    \item \textbf{Realistic}: Performance issues are modeled after real-world patterns (cache misses, algorithmic inefficiency, suboptimal data layout, missing SIMD vectorization, etc.).
    \item \textbf{Measurable}: Each task includes a benchmark driver that quantifies throughput or latency, and a test script that validates both correctness and performance improvement.
    \item \textbf{Diverse}: Tasks span 8 domains including databases, numerical computing, ML runtimes, simulation, text processing, and graph analytics.
    \item \textbf{Cross-platform}: Build and test scripts are provided for both Linux and Windows environments.
\end{itemize}

\subsection{Evaluation Protocol}

An agent is given the \texttt{issue.md} and full access to the \texttt{Code/} directory. The agent must:
\begin{enumerate}
    \item Diagnose the root cause of the performance issue.
    \item Modify the source code to fix the issue.
    \item Ensure the fix passes the automated test script (\texttt{test.sh}), which checks both correctness (no regression) and performance improvement (meets a predefined speedup threshold).
\end{enumerate}

\subsection{Statistics}

\begin{itemize}
    \item Total tasks: 12
    \item Total source files: 272 (C source + headers)
    \item Total lines of code: $\sim$668K
    \item Median task size: $\sim$15K LoC
    \item Language: C (compiled with GCC/MSVC)
    \item Domains covered: 8 (Database, Numerical, Data Structures, ML/DL, Storage, Simulation, Text Processing, Graph Analytics)
\end{itemize}

\subsection{Licensing and Access}

The dataset is publicly available at \url{https://anonymous.4open.science/r/Dataset-D3CC}. All code is provided for research purposes.

\section{Potential Risk}

We see no risk to human subjects or sensitive data: all tasks operate on
self-contained C codebases adapted from open-source projects. The main
residual risks are (i) benchmark gaming, where agents specialize to the
released tasks rather than learning general optimization, which we mitigate
with hidden correctness tests and trajectory audit; (ii) potential misuse
of the documented shortcut techniques, which are standard and already known
in the performance-engineering community; and (iii) the compute cost of
running long-horizon agentic evaluations.

\section{AI Assistants Usage Declaration}

We used AI assistants (large language models) to help draft and polish text
and to assist in analyzing experimental logs. All experimental design,
execution, verification, and final claims are the authors' responsibility,
and the authors reviewed and verified all AI-assisted content.

\end{document}